# Local Energy Landscape in a Simple Liquid

T. Iwashita[1] and T. Egami[1,2,3]

[1]*Joint Institute for Neutron Sciences and Department of Physics and Astronomy, University of Tennessee, Knoxville, TN 37996*

[2]*Department of Materials Science and Engineering, University of Tennessee, Knoxville, TN 37996*

[3]*Oak Ridge National Laboratory, Oak Ridge, TN 37831 USA*

**Abstract**

It is difficult to relate the properties of liquids and glasses directly to their structure because of complexity in the structure which defies precise definition. The potential energy landscape (PEL) approach is a very insightful way to conceptualize the structure-property relationship in liquids and glasses, particularly on the effect of temperature and history. However, because of the highly multi-dimensional nature of the PEL it is hard to determine, or even visualize, the actual details of the energy landscape. In this article we introduce a modified concept of the local energy landscape (LEL) which is limited in phase space, and demonstrate its usefulness using molecular dynamics simulation on a simple liquid at high temperatures. The local energy landscape is given as a function of the local coordination number, the number of the nearest neighbor atoms. The excitations in the LEL corresponds to the so-called $\beta$-relaxation process. The LEL offers a simple but useful starting point to discuss complex phenomena in liquids and glasses.

# 1. Introduction

The potential energy landscape (PEL) concept has been shown to be a powerful tool to study the thermodynamics of complex systems including liquids, glasses, molecules and clusters [1-4]. Although the general properties of PEL are beginning to emerge through various simulations [5-7], it is difficult to visualize PEL directly, particularly because it is so highly multi-dimensional. Consequently the picture of landscape is often schematically drawn by hand [3, 7], and the question what the choice of the horizontal axis to represent atomic configuration should be is still being debated. In this article we demonstrate that the local energy landscape (LEL) in high-temperature liquids can be explicitly presented as a function of the coordination number, and local dynamics can be directly calculated from the LEL. It should be pointed out that even though the global PEL is highly multi-dimensional, in practice there is little need of knowing the full landscape, because vast portions of it are virtually inaccessible due either to its high potential energy or to its small phase space. In order to apply the concept of PEL more effectively to liquids and glasses, high-temperature liquids in particular, we propose to consider a configurationally averaged LEL in easily accessible phase space, instead of the full energy landscape, by using the local coordination number as the axis for atomic configuration. LEL is closely related to the local PEL as discussed below.

In crystalline solids phonons are the elementary excitations of lattice dynamics [8]. But in liquids phonons are strongly scattered and short-lived. This is because the dynamical (Hessian) matrix, of which diagonalization defines the phonons, itself is time-dependent, so that there is no real eigen-state [9]. Instead, we found recently that the local configurational excitation (LCE), which is an action of changing the local topology of atomic connectivity, is the elementary excitation in the liquid, instead of phonons [10]. In Ref. 10 the lifetime of local atomic connectivity, the time to lose or gain one nearest neighbor, $\tau_{LC}$, was found to be equal to the Maxwell relaxation time, $\tau_M = \eta/G_\infty$, where $\eta$ is shear viscosity and $G_\infty$ is the high-frequency shear modulus, above the crossover temperature, $T_A$. This is an important finding, because this result connects a microscopic time-scale, $\tau_{LC}$, directly to a macroscopic time-scale of liquid dynamics, $\tau_M$.

In the present work through molecular dynamics (MD) simulation of a simple liquid, a metallic Cu-Zr alloy liquid, we show that the local dynamics of the LCE can be described very well in terms of the LEL and excitations within the LEL. In the language of the PEL theory a high temperature liquid system is supposed to fly above the PEL as in a gaseous state. However, a high temperature liquid is a condensed matter with strong atomic correlations, not a gas. On the other hand because of localization of atomic dynamics in high temperature liquids [10] all atoms of the same specie see the same local PEL. We propose that a better description is that the global PEL is reduced to the local atomic-level energy landscape in high-temperature liquids, because the system chooses the states with the highest degeneracy to maximize entropy.

In the PEL approach the state of the system is characterized by its inherent structure obtained by quenching the system to $T = 0$ [3, 5], and the dynamics of the system is studied by its migration from one inherent state to the other [7]. Myopic details of the state are not of concern. Whereas this far-sighted approach renders the method power to investigate extremely complex systems, physical intuition suffers. In the present work we provide explicit depiction of the microscopic state in terms of the local coordination number, focusing on high-temperature liquids for which the PEL is particularly simple. Extension to supercooled liquids and glasses will be discussed.

## 2. Molecular dynamics simulation

The three-dimensional MD simulations on $Cu_{56}Zr_{44}$ were performed using LAMMPS on a model in a cubic box under periodic boundary condition. Our system consists of 16,000 atoms under the NVT ensemble at a number density of 0.05864 Å$^{-3}$, and for the interaction between atoms we use the embedded atom method (EAM) potential [11]. The temperatures studied range from 850K to 5000K, all of which are above $T_g$ (= 700 K). The value of $T_g$ was obtained from a jump in specific heat during a cooling process at $5 \times 10^{10}$ K/sec. The model has a crossover temperature $T_A$ below which viscosity increases rapidly in a super-Arrhenius fashion, which is about 1600K [10]. Equilibrium simulations allow for the calculation of thermodynamic or transport properties, such as energy and viscosity, and in the well-equilibrated liquid states the transition rates at which an atom loses or gains one neighbor were calculated as we discuss later. To improve the statistics of the results we averaged over 10 independent simulation runs. In the simulation, time and length are expressed in units of ps and Å, respectively, and the MD time steps were 0.001 ps for temperatures below 3000K and 0.0005 ps for temperatures above 3000K.

## 3. Local coordination number

Defining the structure by atomic connectivity is natural for covalent glasses such as $B_2O_3$, which is well described by the continuous random network (CRN) model [12]. However, even though strong covalent atomic bonds do not exist in metallic systems, one can still define the nearest neighbors by the first peak in the atomic pair-density function (PDF), using the first minimum in the PDF, $r_{min}$, as the cut-off; if the interatomic distance between two atoms is shorter than $r_{min}$ they are defined as nearest neighbors. The number of nearest neighbors is called the coordination number, $N_C$, and varies from an atom to another. The topology of atomic connectivity network thus defined is an effective way to characterize the structure of metallic liquids and glasses as the CRN structure with loose restriction on the coordination number. For a binary alloy $Cu_{56}Zr_{44}$ the first minimum in partial PDFs was used as the cut-off value. The cut-off values for Cu-Cu, Cu-Zr, and Zr-Zr pairs are 3.172 Å, 3.692 Å, and 4.108 Å, respectively. Alternatively it is possible to define the nearest neighbors through the Voronoi construction [13]. However, Voronoi method tends to overestimate $N_C$ by counting the neighbors with small Voronoi face.

The distribution of the value of $N_C(\alpha)$, local coordination number for the $\alpha$ atom where $\alpha$ = Cu or Zr, $P_\alpha(N_C(\alpha))$, is given in Fig. 1 for the model liquid $Cu_{56}Zr_{44}$. Here the distribution is given by $P_\alpha(N_c) = n_\alpha(N_c)/N$ where $n_\alpha(N_c)$ is the number of $\alpha$ atoms having $N_c$ nearest neighbors and $N$ is the total number of atoms. The probability is normalized by,

$$\sum_{Nc}\left[P_{Cu}(Nc) + P_{Zr}(Nc)\right] = 1. \tag{1}$$

The average coordination number is a weak dependence on $T$, and $\langle N_c(Zr)\rangle$ is about 15 and $\langle N_c(Cu)\rangle$ is about 10 ~ 11, reflecting their atomic sizes ($r_{Zr}$ = 1.52 Å, $r_{Cu}$ = 1.32 Å [11]). The distribution is nearly Gaussian, and the peak of the distribution shifts to a lower number and becomes wider as temperature is raised. This distribution function can be expressed in terms of the local effective free energy, $E^\alpha(N_C(\alpha))$, by

$$P_\alpha(N_C(\alpha)) = P_{\alpha,0}\exp\left(-\frac{E^\alpha(N_C(\alpha))}{k_B T}\right). \tag{2}$$

$E^\alpha(N_C(\alpha))$ can be directly calculated using eq. (2) and is shown for Zr and Cu in Fig. 2. Now the deviation in $N_C$ from the thermal average, $N_C(\alpha) - \langle N_C(\alpha)\rangle$, is proportional to the atomic-level pressure, $p_\alpha$ [14]. For a monoatomic system it is given by

$$p = \frac{6\sqrt{3}-9}{4\pi} B\left(N_C - \langle N_C \rangle\right), \tag{3}$$

where $B$ is the bulk modulus [15]. The local elastic self-energy can be expressed as $V\langle p^2\rangle/2B$, where $V$ is the atomic volume and $\langle p^2\rangle$ is the second moment of distributed atomic-level pressure [14,16]. Furthermore, we have shown earlier that $p$ obeys the equipartition law at high temperatures [16, 17]. Therefore $E(N_C)$ can be explicitly expressed as [15]

$$E(N_C) = VB \frac{27(7-4\sqrt{3})}{16\pi^2} \left(N_C - \langle N_C \rangle\right)^2. \tag{4}$$

The extension to the case of a binary alloy is given in Appendix. Indeed the data fit this expression well as shown in Fig. 2 at 3000K in spite of various approximations introduced in deriving eq. (4). At high temperatures the higher-order terms become important, partly because the volume is kept constant. Now the fact that the equipartition law is obeyed means that $N_C$ is effectively an independent local variable in a liquid. Therefore even though the coordination number is an integer, we can generalize it as a continuous variable and use it as the coordinate for the LEL. The integral values of $N_C$ correspond to the minima in the LEL, expressed by eq. (2). In-between the integral values of $N_C$ the LEL has energy barriers of which heights determine the transition rates between neighboring $N_C$s.

## 4. Transition rates for coordination number

As time goes forward an atom may lose one of the nearest neighbors which moves on to become the second neighbor. So the local coordination number of this atom, $N_C(i)$, is reduced by one. This happens simultaneously to two atoms, $i$ and $j$, when the connectivity between $i$ and $j$ is lost. This justifies the description of this action as "breaking of a bond". An atom can also gain a new nearest neighbor. Then a "new bond" is created, and the local coordination number is increased by one.

Figure 3 shows typical examples of the time evolution of $N_C$ for several atoms. The flow chart for the dynamical process of $N_C$, $(b_i, f_i)$, is displayed in Fig. 4. Here $b_i$ and $f_i$ are two connectivity parameters, which measure how many bonds are broken or formed with respect to a reference state for the atom $i$. Let us consider that at $t = 0$ an atom is in an $N_C$-coordinated state, which serves as the reference state denoted by $(b_i, f_i) = (0, 0)$. As time goes by the atom loses or gains one neighbor at $t = t_1$, and the states of the atom are given by $(1, 0)$ or $(0, 1)$. Hereafter, the two pathways through which the atom takes $(1, 0)$ or $(0, 1)$ are referred to as process I or process II, respectively.

Next we consider the escape time, $t_2 (> t_1)$, at which the states excited at $t = t_1$ move to different states. As shown in Fig. 4 the possible next states that can be taken at $t = t_2$ are given by $(2, 0)$ or $(1, 1)$ for process I and $(0, 2)$ or $(1, 1)$ for process II. Often fluctuations let the atom go back to the previous states, $(1, 0)$ or $(0, 1)$. For this case the value of $t_2$ remains unassigned until the atom again reaches one of the states, $(2, 0)$, $(1, 1)$ and $(0, 2)$. In order to differentiate if the state arrived at $t = t_2$ is due to fluctuations or structural changes, we need to know further transition states, $(3, 0)$, $(2, 1)$, $(1, 2)$, or $(0, 1)$ for process I and $(0, 3)$, $(2, 1)$, $(1, 2)$ or $(1, 0)$ for process II. When the atom takes one of such states at $t = t_3 (> t_2)$, we assign $t_2$, just before $t_3$, as the time at which the atom eventually escapes from the states with $(1,0)$ or $(0,1)$. Within a time interval of $[t_1, t_2]$, some atoms were found to fluctuate back and forth between the two states, especially at high temperatures. In such a case it is difficult to identify exactly when the excitations take place. For the present purpose, it is more convenient to define the time during which the atom lies in the states with (1, 0) or (0, 1) for $t \in [t_1, t_2]$ as $\tau_{LC}$. This can approximate the lifetime of the states with (1, 0) or (0, 1). If the atom reaches a state with $N_C$ at $t = t_1$ and then undergoes a transition from this state to a state with $N_C \pm 1$ at $t = t_2$, $\tau_{LC} = t_2 - t_1$ is expressed as $\tau_{LC}(N_c|N_c \pm 1)$. Also we can count the number of atoms undergoing a transition from $N_c$ coordinated state to a $N_c \pm 1$ coordinated state in unit time, which is denoted by $\Delta n(N_c|N_c \pm 1)$. In the present analysis the probability that bond breaking and formation occurs simultaneously on the same atom is small, and such atoms (less than 6% of atoms) were neglected. Note that the value of $t_1$ was reset only when the atom goes from (0, 1) to (1, 0) via (0, 0) or vice versa.

Using, $\tau_{LC}^\alpha(N_C|N_C+1)$, $\Delta n^\alpha(N_C|N_C+1)$ and $n^\alpha(N_C)$ we define the rate at which transitions from $N_c$ to $N_c \pm 1$ occurs by

$$k_{LC}^\alpha(N_C|N_C \pm 1) = \frac{\Delta n^\alpha(N_C|N_C \pm 1)}{n^\alpha(N_C)\tau_{LC}^\alpha(N_C|N_C \pm 1)}, \quad (5)$$

and $n^\alpha(N_C) = \Delta n^\alpha(N_C|N_C+1) + \Delta n^\alpha(N_C|N_C-1)$. As shown in Fig. 5 $k_{LC}^\alpha$ strongly depends on temperature as well as on the local coordination number, $N_C$. This is reasonable because it is easier to lose a neighbor when $N_C$ is large, and vice versa. The combined transition rates,

$$k_{LC}(\alpha) = (1/2)\left[k_{LC}^\alpha(N_C|N_C-1) + k_{LC}^\alpha(N_C|N_C+1)\right], \quad (6)$$

are shown in Fig. 6. They show when $N_C$ is close to its average, $<N_C>$, the transition rates are low so that the system is stable as expected, whereas it is unstable when it is far away.

## 5. Temperature dependence of the transition rate

It is found that $k_{LC}^{\alpha}$ has the Arrhenian temperature dependence over a wide temperature range as shown in Fig. 7, except for $N_C = 12$ for Cu which shows an anomalous behavior. This exceptional behavior will be discussed later. From this result we can determine the activation energy for the process $N_C \to N_C - 1$, $\Delta E^{\alpha}(N_C|N_C - 1)$ and that for $N_C \to N_C + 1$, $\Delta E^{\alpha}(N_C|N_C + 1)$, and calculated it as $\Delta E^{\alpha} = -k_B T \log(k_{LC}^{\alpha}(T)/k_{LC}^{\alpha}(\infty))$ for each process shown in Fig.7.

By combining these activation energies with $E^{\alpha}(N_C)$ in eq. (2), which represents the energy of the local minimum in the LEL for each coordination state, we can construct the local energy landscape as a function of $N_C$ as shown in Fig. 8. The saddle point energies in the LEL are given by $E_{saddle}^{\alpha+}(N_c + 1/2) = E^{\alpha}(N_c) + \Delta E^{\alpha}(N_c|N_c + 1)$ for the process $N_C \to N_C + 1$ and by $E_{saddle}^{\alpha-}(N_c + 1/2) = E^{\alpha}(N_c + 1) + \Delta E^{\alpha}(N_c + 1|N_c)$ for the process $N_C \to N_C - 1$. Here it is seen that the salient part of the LEL changes little with temperature. The portions of the LEL with high values of $N_C$ increase with temperature. The state with high coordination number involves the nearest neighbor atoms which are closer to each other. At high temperatures this results in accessing the more strongly repulsive part of the interatomic potential, increasing the local potential energy.

The saddle point energies for the activation processes $N_C \to N_C - 1$ and $N_C \to N_C + 1$ are not exactly identical. But these small differences must be due to statistical noise, because the law of detailed balance requires them to be equal. The LEL shown in Fig. 9 is the average of the two LEL's for the processes to increase and decrease $N_C$. Note that this landscape is different from the full energy landscape in that here all the variables other than $N_C$ are averaged out in statistical sampling, and in doing so only the portion of the phase space accessible at each temperature is considered. It clearly shows that the coordination states close to $<N_C>$ are more stable, and the activation processes to change $N_C$, LCEs, are the elementary process of activation.

## 6. Comparison with viscosity

The total LCE transition rates of the system,

$$k_{LC}^{\pm} = \sum_{N_C} k_{LC}^{Cu}(N_C|N_C \pm 1) P_{Cu}(N_C) + \sum_{N_C} k_{LC}^{Zr}(N_C|N_C \pm 1) P_{Zr}(N_C) , \tag{7}$$

are plotted against $1/T$ in Fig. 10, and are compared to the Maxwell relaxation time of the system calculated using viscosity obtained by the Green-Kubo formula [10]. Obviously $k_{LC}^{+} = k_{LC}^{-}$ in order to preserve the steady state. As shown in Ref. 10 $k_{LC} = 1/\tau_M$ above the crossover temperature, $T_A \sim 1600$ K. It is observed at least within the temperature range studied the temperature dependence of $k_{LC}^{\pm}$ remains Arrhenian down to well below $T_A$, even though the Maxwell relaxation time, thus viscosity, deviates from the Arrhenian behavior below $T_A$. This result supports the earlier conjecture that the nature of the LEL and the excitation to change the local coordination number, LCE, is basically independent of temperature, justifying the definition of LCE as the elementary excitation in liquids. However, the interactions among the LCE's change as the system is cooled through $T_A$. Above $T_A$ the excitations are independent of each other because the Maxwell relaxation time is shorter than the time for the phonon to propagate over one atomic distance, but below $T_A$ they start to interact through exchanging phonons [10]. Note that above $T_A$ $1/k_{LC}$ calculated from eq. (7) gives almost the same value as $\tau_{LC}$ defined in Ref. 10.

In terms of the relaxation phenomena it is most likely that the LCE is related to the so-called $\beta$-relaxation [18, 19], whereas the Maxwell relaxation represents the macroscopic viscosity. To prove this point we show the imaginary part of the dynamic shear modulus, $G''(\omega)$, at $T = 3000$ K ($> T_A$) and 850 K ($< T_A$) in Fig. 11. The dynamic shear moduli were defined as the Fourier transform of stress-stress correlation function with respect to time:

$$G'(\omega) + iG''(\omega) = i\omega \int dt \langle \sigma_{xy}(t)\sigma_{xy}(0)\rangle \exp(i\omega t) \tag{8}$$

where $G'(\omega)$ and $G''(\omega)$ are called storage modulus and loss modulus, respectively, and $\sigma_{xy}$ is the shear component of macroscopic stress tensor and $\omega$ is angular frequency. At $T = 3000$ K $G''(\omega)$ has only one peak at the frequency $\omega_p \sim \omega_{LC} = 1/\tau_{LC} = 1/\tau_M$, as expected from the Debye relaxation phenomena. At $T = 850$ K $G''(\omega)$ shows two peaks, the fast dynamic peak due to phonons and the other relaxation peak tracking $1/\tau_M$. However, there is additional weight between the two peaks, usually attributed to the extended $\beta$-relaxation wing [20]. $\omega_{LC}$ falls right at the extended $\beta$-relaxation wing, suggesting that the LCE or combination of the LCEs could be the origin of the $\beta$-relaxation process.

It has been known by the PEL approach that the local excitations out of the metabasin have Arrhenian temperature dependence [7, 21]. However, the precise nature of such excitations is not known, except that they are thought to be related to breaking of the cage. In this work we explicitly identified the excitations as cutting or forming of atomic bonds, represented by changing the coordination number

## 7. Discussion

Whereas the PEL describes the energy states of the inherent structures [x] the LEL is determined by the population of the $N_C$ states and the transition rates among them. Therefore the LEL reflects the local free energy landscape rather than the local potential landscape. However, the difference may not be significant. As shown in Fig. 2 the energies of the minima are expressed by eq. (4) in terms of the elastic self-energy. Deviations at the both ends most likely reflect anharmonicity than the entropical effect. Also as shown in Fig. 7 the activation energy does not change much with temperature, again suggesting that the entropical effects are minimal. Therefore we believe that the LEL reflects the local PEL rather faithfully.

The result shown in Fig. 9 suggests that the LEL of a liquid at high temperatures can be expressed as

$$E^{\alpha}_{PEL} = C^{\alpha}\left(N_C(\alpha) - \langle N_C(\alpha) \rangle\right)^2 + \beta_{\alpha} \sin^2\left(\pi N_C(\alpha)\right) + HO\left(N_C(\alpha) - \langle N_C(\alpha) \rangle\right) \quad (8)$$

where $C^{\alpha}$ is given by eq. (4) for a monoatomic system and in Appendix for a binary alloy. The higher order term, $HO$, due to anharmonicity increases with temperature. The value of $\beta_{\alpha}$ appears to be related to the bond energy [15]. It depends only weakly on $N_C$, so in eq. (8) we neglected this dependence. This expression provides a concrete and simple picture of the LEL of a liquid, and allows various properties to be calculated. We showed an example of viscosity in Fig. 10.

As shown in Fig. 6 the lowest transition rate for Cu is at $N_C = 10$ at high temperatures, but the transition rate for $N_C = 12$ becomes unusually low at low temperatures. As shown in Fig. 7 the transition rate for $N_C = 12$ significantly deviates downward from the Arrhenian behavior below 1100 K. Consequently the rate for $N_C = 13$ deviates upward. This anomalous behavior must be a consequence of the formation of stable icosahedral clusters. Icosahedral clusters are often found to dominate at low temperatures [11, 22-24]. As pointed out by Frank long time ago [25] icosahedral clusters are not conducive to crystallization, and contribute to enhancing supercoiling. Thus there have been many discussions to relate the formation of icosahedral clusters to the stability of the glassy state [11, 22-27]. However, the total transition rates shown in Fig. 9 do not show significant deviation from the Arrhenius behavior. Therefore whereas the formation of icosahedral clusters is certainly favored, it does not appear to be the main driver for glass formation through the rapid increase in viscosity. As Frank originally suggested their main contribution to glass stability could be to increase the boundary energy between the liquid and the crystal and slow down the crystallization kinetics. It is also most likely that icosahedra are found stable mainly because the coordination 12 is often close to the average, $<N_C>$, because the ideal coordination in a mono-atomic system is $4\pi$ (= 12.56) [28]. Indeed for Zr ($N_C$ =) 12 is far from the average, and the state with $N_C = 12$ is not preferred.

On the other hand, the LEL clearly shows the emergence of a highly degenerate state at low temperatures. For instance, for Cu, states with $N_C = 10$, 11 and 12 are almost equally preferred at low temperatures. For Zr the state with $N_C = 15$ is at the energy minimum, but the states with $N_C = 14$ and 16 are populated as well. Therefore we can summarize the features of the LEL in relation to glass stability as follows;

1. Atoms with $N_C$ close to the average, $<N_C>$, are stable.
2. At low temperatures atoms choose multiple states, not one state with a particular $N_C$, but states with a range of $N_C$.
3. Such degeneracy is at the core of glass formation, because the preference of a single $N_C$ tends to drive the system to the crystalline state.
4. Formation of exceptionally stable clusters, such as the icosahedral clusters, contributes to the glass stability, but they are not the main factors.
5. Exceptionally stable clusters are preferred, however, only when the condition 1 is met. For instance, icosahedra ($N_C = 12$) are stable, only when $<N_C>$ is close to 12.

The atoms with $N_C$ close to the average have small atomic-level pressure, because the atomic-level pressure is proportional to $N_C - <N_C>$ as in eq. (3). The results in Figs. 8 and 9 justify the idea that the atoms with small values of pressure, either negative or positive, are solid-like, and those with large pressure, thus the atoms with $N_C$ far from the average, are unstable and liquid-like [29].

As shown in Ref. 10 LCEs are independent at high temperatures above $T_A$, but interact with each other through the exchange of phonons. It has been known that the atomic-level stresses are spatially correlated below $T_A$ [16]. The Eshelby theory, which is a continuum theory of elastic inclusion [30], is known to describe the dressing of the atomic-level stresses by the long-range elastic field [14, 29]. Therefore the interaction between the LCEs may be described in terms of the Eshelby theory. Indeed we found recently that the correlations among the atomic-level stresses follow the symmetry of the Eshelby field as we will discuss elsewhere. It is possible that the interaction among LCEs through the long-range Eshelby field leads to super-Arrhenius behavior of viscosity below $T_A$.

In this work we employed the NVT ensemble rather than the NPT ensemble. In the NPT ensemble the increase in temperature results in thermal expansion, and eventually in evaporation. Such significant changes in the structure would make the activation energy strongly temperature dependent, and would not allow it to be determined from the Arrhenius relation as was done in this work. On the other hand the NVT ensemble provides the liquid structure much less dependent on temperature. For this reason we decided to use the NVT condition.

Finally it is interesting to speculate if the LEL concept could be extended to other classes of liquids and glasses, such as the network-forming glass-formers where there exist well-defined bonds between atoms (covalent bonds, hydrogen bonds, etc). Strong glasses, such as silicates and borates, are characterized by fixed coordination numbers. However, at elevated temperatures deviations from the fixed environment start to occur through bond breaking, and they determine the properties such as viscosity. Therefore it is likely that the same analysis would work, and the difference is merely the enhanced energy scale for the LEL, both the height of the barriers as well as the energy states. On the other hand in more complex glasses, such as soda glasses and polymers, strong covalent bonds and weak ionic or Van-der-Waals bonds coexist. In such a case the LEL would have two subsystems, the LEL in small energy scales for weak bonds and the LEL with large energy scale for covalent bonds. In either case, the extension of the present approach may well prove to be very useful.

## 8. Conclusions

In this work we have demonstrated that the local energy landscape (LEL) in high temperature liquids can be explicitly obtained based upon the information on the distribution of the local coordination number $N_C$ and the rates to change it. The results of simulation on liquid Cu-Zr give an intuitive picture of the relative stability and dynamics of each coordination state. The minima in the LEL are characterized by the coordination number. Atoms with $N_C$ closest to the average, $<N_C>$, have the highest stability. The system makes thermally activated jumps from one coordination state to another over the barrier and these jumps provide the basis for the $\beta$-relaxation. As a liquid is cooled the LEL remains largely unchanged, except for the case of $N_C = 12$ for Cu, which is due to formation of icosahedral clusters. As shown here the LEL provides intuitive and yet realistic understanding of the local structure and complex dynamics of normal and supercooled liquids.

**Acknowledgments**

The authors are grateful to S. Sastry, S. Yip, J. S. Langer, Y. Q. Chen, K. Kelton and J. R. Morris for useful discussions. The work was supported by the Department of Energy, Office of Sciences, Basic Energy Sciences, Materials Science and Engineering Division.

**Appendix : Predicting the probability function for Nc in a binary alloy**

Based on local elastic self-energy we derive an expression for local effective energy, given in Eq. (2) as a function of coordination number in a binary alloy $Cu_{56}Zr_{44}$. Let us consider a monoatomic system composed of spheres with an effective radius,

$$r_{ave} = \sqrt{f_{cu}r_{cu}^2 + f_{zr}r_{zr}^2} \qquad (A.1)$$

where $f_{cu} = 0.56$ and $f_{zr} = 0.44$. The packing problem of a single atom with a radius of $r$ embedded in the monoatomic system of atoms with $r_{ave}$ gives an approximate expression for the coordination number for the embedded atom [28]

$$N_c(x) = 4\pi \left(1 - \frac{\sqrt{3}}{2}\right)(1 + x)\left(1 + x + \sqrt{x(x+2)}\right) \qquad (A.2)$$

where the size ratio $x = r/r_{ave}$. The validity of Eq. (A. 2) was tested numerically, and it was shown that the results predicted by Eq. (A. 2) agree very well with simulation results [28]. Similar analysis based on Eq. (A.2) also allows us to calculate coordination number in a binary glass with different size ratio [31].

The atomic level volume strain is given by

$$\epsilon_V = \frac{3}{2}\frac{\Delta x}{x} = \frac{3}{2}\frac{1}{x}\left(\frac{dx}{dN_c(x)}\right)_{x=x_\alpha} \Delta N_C \qquad (A.3)$$

where $x_\alpha = r_\alpha/r_{ave}$ and $r_\alpha$ is the radius of embedded $\alpha$ atom ($\alpha$ = Cu or Zr). Note that the atomic level strain is different from the volume strain obtained from uniform volume expansion by a factor of two, because the atomic volume strain is defined as the change in the pair distance between neighboring atoms.

Using Eq. (A.2) and (A.3) the atomic level pressure of the embedded atom, $p_\alpha$, is expressed as

$$p_\alpha = B_\alpha \epsilon_V = \frac{3B_\alpha}{2x} \left(\frac{dx}{dN_c(x)}\right)_{x=x_\alpha} \Delta N_c(x). \tag{A.4}$$

where $B_\alpha$ is the bulk modulus for the embedded atom. The local elastic self-energy is then given by

$$E_p^\alpha = \frac{p_\alpha^2 V_\alpha}{2B_\alpha} = \frac{9B_\alpha V_\alpha}{8x^2} \left(\frac{dx}{dN_c(x)}\right)^2_{x=x_\alpha} \left(\Delta N_c(x)\right)^2. \tag{A.5}$$

where $V_\alpha$ is the atomic volume for the $\alpha$ atom. From the equipartition theorem for the fluctuations in $p_\alpha$[15], the local elastic energy $E_p^\alpha$ is related to $C^\alpha$ in Eq. (8) by

$$C^\alpha = 2E_p^\alpha. \tag{A.6}$$

Therefore the final form for local effective energy in Eq. (2) can be expressed as

$$E_\alpha(N_c) = 2E_p^\alpha(N_c - \langle N_c \rangle)^2 \tag{A.7}$$

The curves thus predicted by (A.7) are shown in Fig.2, and the good agreement with simulation results suggests that the distribution of *Nc* is determined largely by local elastic energy.

**Figure captions:**

Figure 1. Distribution of the coordination number, $N_C$, at various temperatures for liquid $Cu_{56}Zr_{44}$ for Cu and Zr. The lines are Gaussian fits to the data for each temperature.

Figure 2. The energies of the states with various $N_C$, deduced by eq. (2) at 3000K for Cu and Zr. The solid lines represent theoretical predictions based on the local elastic energy given by eq. (A. 7).

Figure 3. Examples of the time evolution of $N_C$ for several atoms The atoms are constantly undergoing a discontinuous change in $N_c$, reflecting the discreteness of atoms.

Figure 4. The flow chart for the dynamical process of $N_C$. In the liquid states the local coordination number of each atom changes with time, and the connectivity parameters, $(b_i, f_i)$, are introduced to describe how the local structure of an atom changes with time. These changes essentially take place in a discontinuous manner. See text (section 3) for definitions.

Figure 5. Coordination number dependence of the transition rates for Cu and Zr. We see that the transition rates strongly depend on $N_c$. For the process to gain a neighbor (Left) it is easy to gain a neighbor when $N_c$ is low, than when $N_c$ is high. Similarly for the process to lose a neighbor (Right) it is more difficult to gain a neighbor when $N_c$ is low. It is also dependent on temperature, because the atomic mobility increases with temperature.

Figure 6. The combined rate of local configurational change, $k_{LC}$, for Cu (left) and Zr (right) as a function of $N_C$ for $Cu_{56}Zr_{44}$ liquid. Here $k_{LC}(Cu) = (1/2)\left[k_{LC}^{Cu}(N_C|N_C-1) + k_{LC}^{Cu}(N_C|N_C+1)\right]$ and $k_{LC}(Zr) = (1/2)\left[k_{LC}^{Zr}(N_C|N_C-1) + k_{LC}^{Zr}(N_C|N_C+1)\right]$. The lower the rate, more stable the atom is. Thus $k_{LC}$ is minimum near the average coordination of each element at high temperatures. For Cu atoms below 900K the atoms with $N_c = 12$ gain stability due to formation of icosahedral clusters, but the atoms with $N_c = 11$ or 13 also gain stability. For Zr atoms at low temperatures the atoms with $N_c = 15$ become most stable

Figure 7. Temperature dependence of the transition rates for Cu and Zr, plotted against $1/T$. For high temperatures above $T_A$ the transition rates exhibit an Arrhenius behavior with a well-defined activation energy. The lines represent the Arrhenius fit with $k = k_\infty \exp(-\Delta E_a / k_B T)$. We see that the Arrhenius activation energy depends on $N_c$ as well as the type of atoms and the process to gain or lose one neighbor. As temperature is decreased the activation energy shows temperature dependence.

Figure 8. The local energy landscape for $Cu_{56}Zr_{44}$ at various temperatures for Cu (above) and Zr (below), for step-down (left) and step-up (right). The local energy minimum at integer values of $N_c$ was calculated by Eq.(2) and the energy barrier at half integer values was given by the activation energy obtained from the Arrhenius fit at high temperatures shown in Fig. 7. Then the saddle point energies at $N_c+1/2$ are given by $E_{saddle}^{\alpha+}(N_c+1/2) = E^\alpha(N_c) + \Delta E^\alpha(N_c|N_c+1)$

for the process to increase $N_c$ and by $E^{\alpha-}_{saddle}(N_c+1/2) = E^{\alpha}(N_c+1) + \Delta E^{\alpha}(N_c+1|N_c)$ for the process to decrease $N_c$.

Figure 9. The local energy landscape for $Cu_{56}Zr_{44}$ at various temperatures for Cu (left) and Zr (right). The local energy landscape was calculated by averaging two LELs for the processes to increase and decrease $N_c$ by one. The energy minimum at integer values was determined by Eq. (2) and the saddle point energy at $N_c+1/2$ was given by $(E^{\alpha+}_{saddle} + E^{\alpha-}_{saddle})/2$.

Figure 10. The compositionally averaged rates of configurational change, $k^{\pm}_{LC}$, compared with the inverse of the Maxwell relaxation time, $\tau_M$. Above $T_A(=1700K)$ $k^{\pm}_{LC} = 1/\tau_M$, while below $T_A$ $k^{\pm}_{LC} > 1/\tau_M$. The Arrhenius activation energy at high temperatures is 0.248 eV.

Figure 11. The imaginary part of the dynamic shear modulus, $G''(\omega)$, which shows the loss spectrum of internal friction. At $T = 3000$ K the spectrum has only one peak, but at $T = 850$ K it splits into two peaks, with extra-weight in between the two peaks, which corresponds to the $\beta$-relaxation wing. $\omega_{LC} = 1/\tau_{LC}$ falls right at the extended $\beta$-relaxation wing, suggesting that the LCE or combination of the LCEs could be the origin of the $\beta$-relaxation process. The solid line represents a Newtonian behavior with $\eta\omega$ and the viscosity $\eta$ was calculated from the Green-Kubo equation for shear stress. The dash line is the high-frequency shear modulus, $G_{\infty}(= V\langle\sigma^2_{xy}\rangle/k_BT)$.

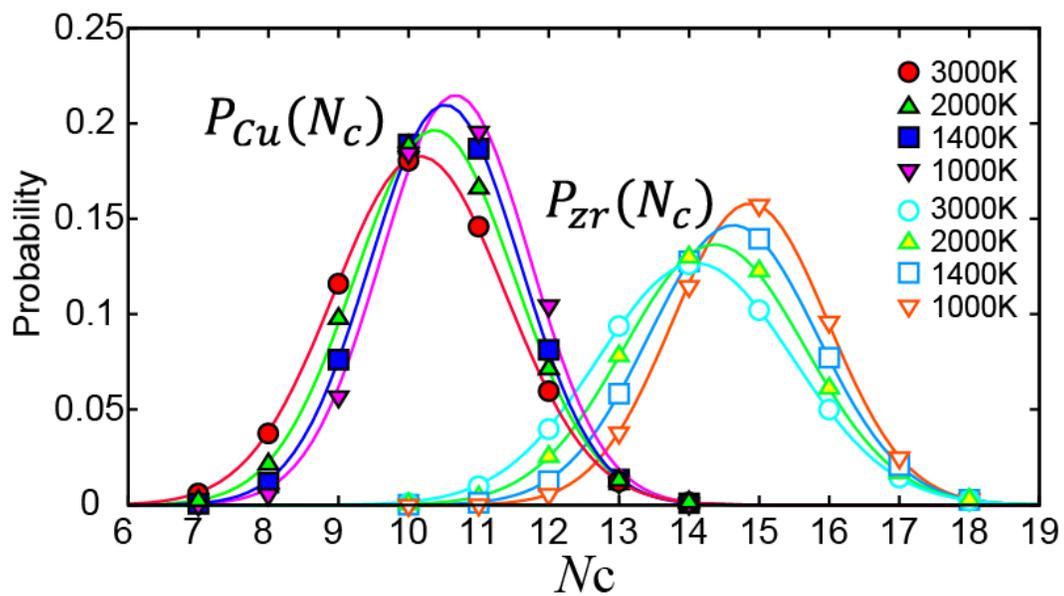

Fig. 1

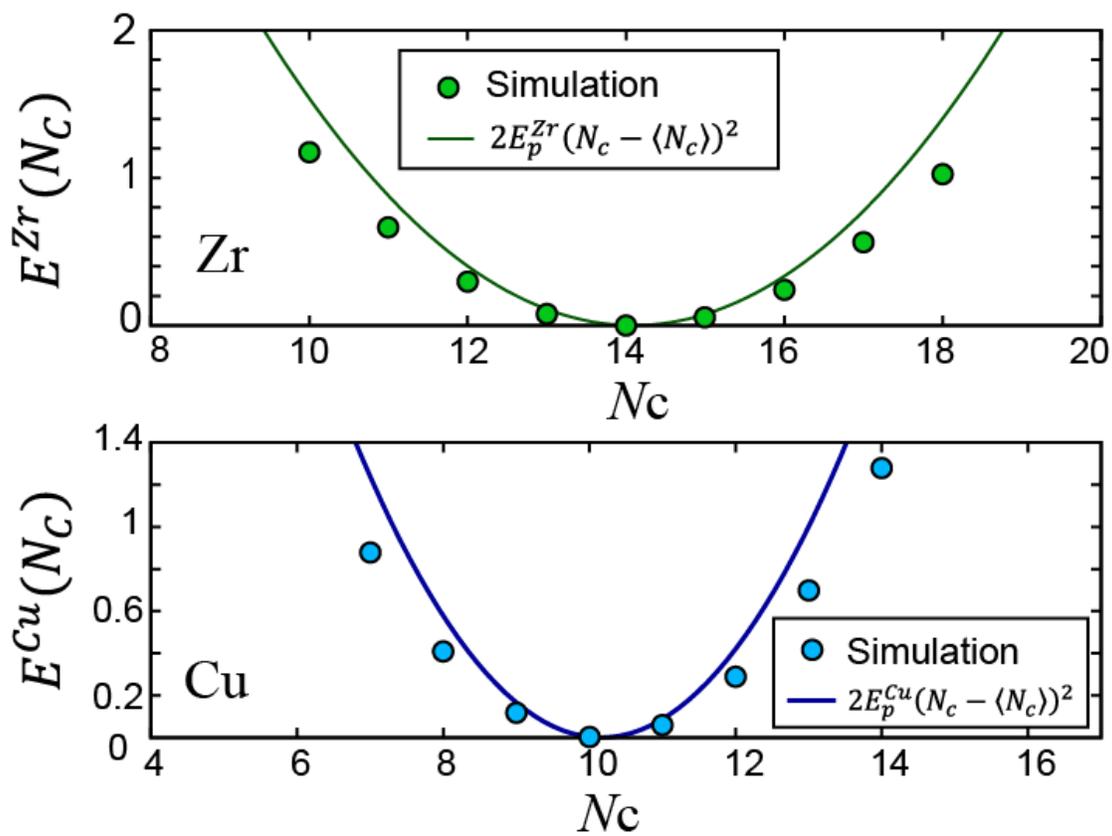

Fig. 2

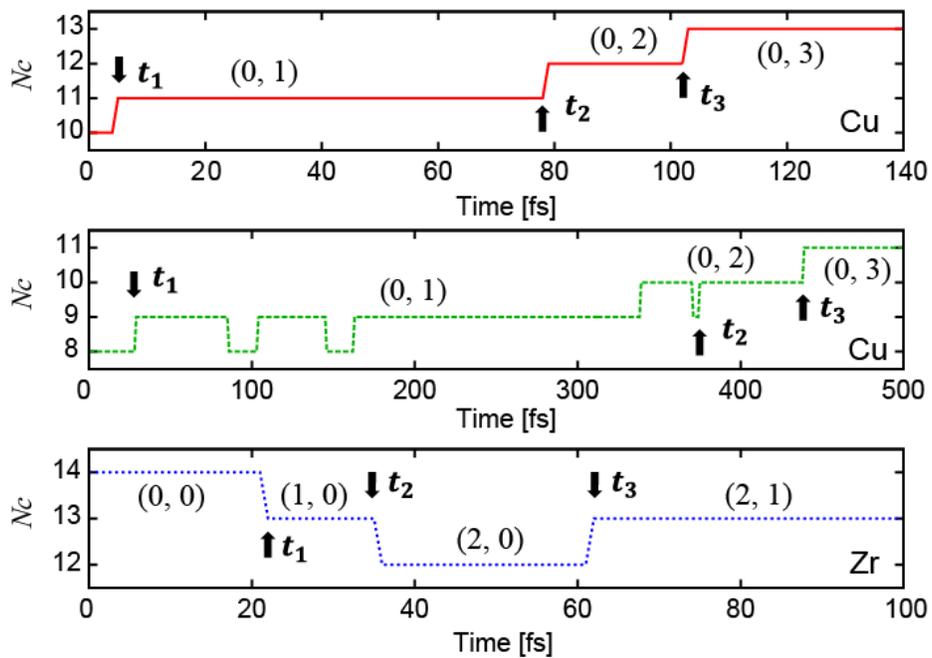

Fig. 3

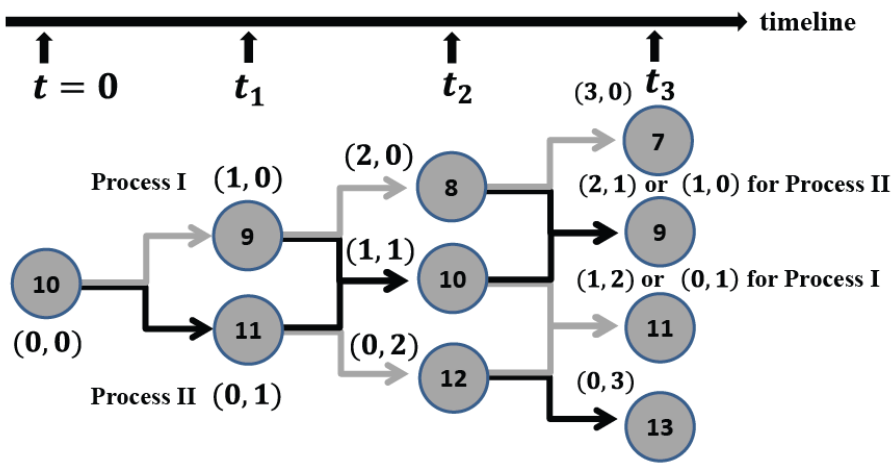

Fig. 4

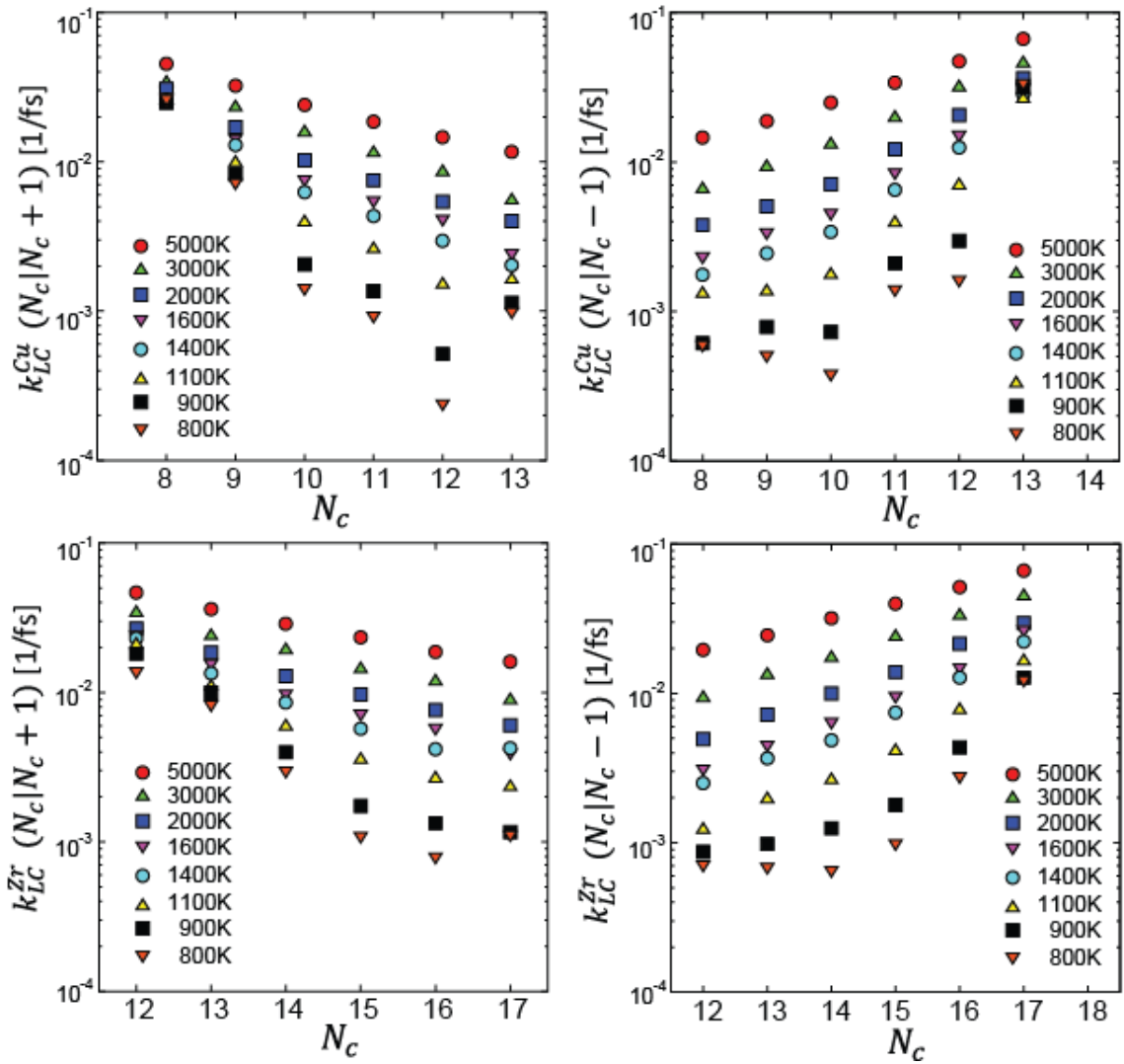

Fig. 5

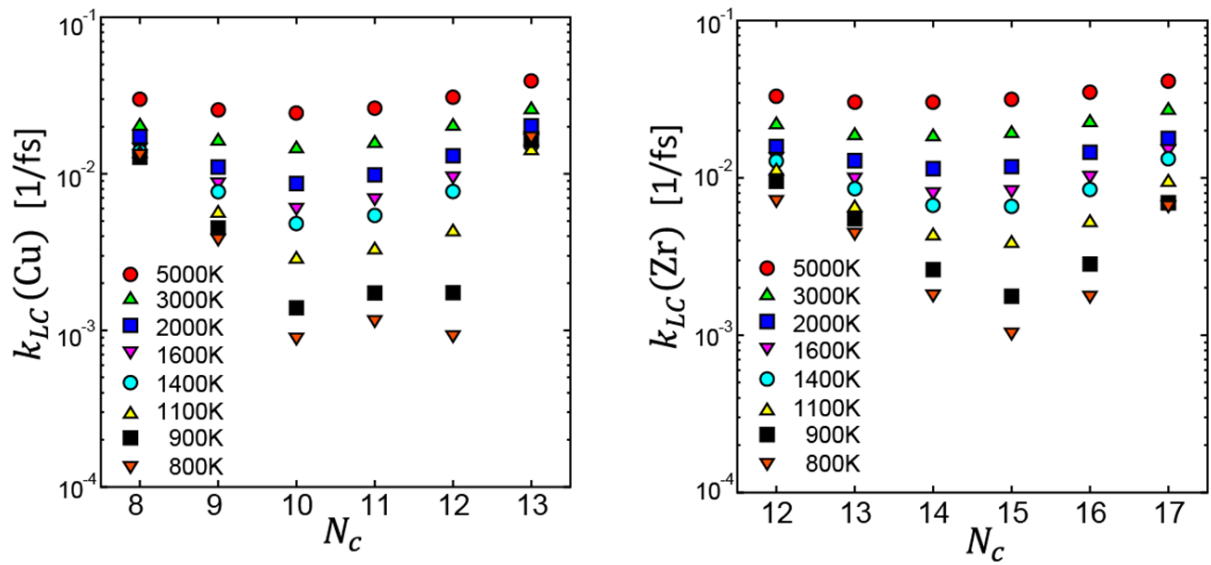

Fig. 6

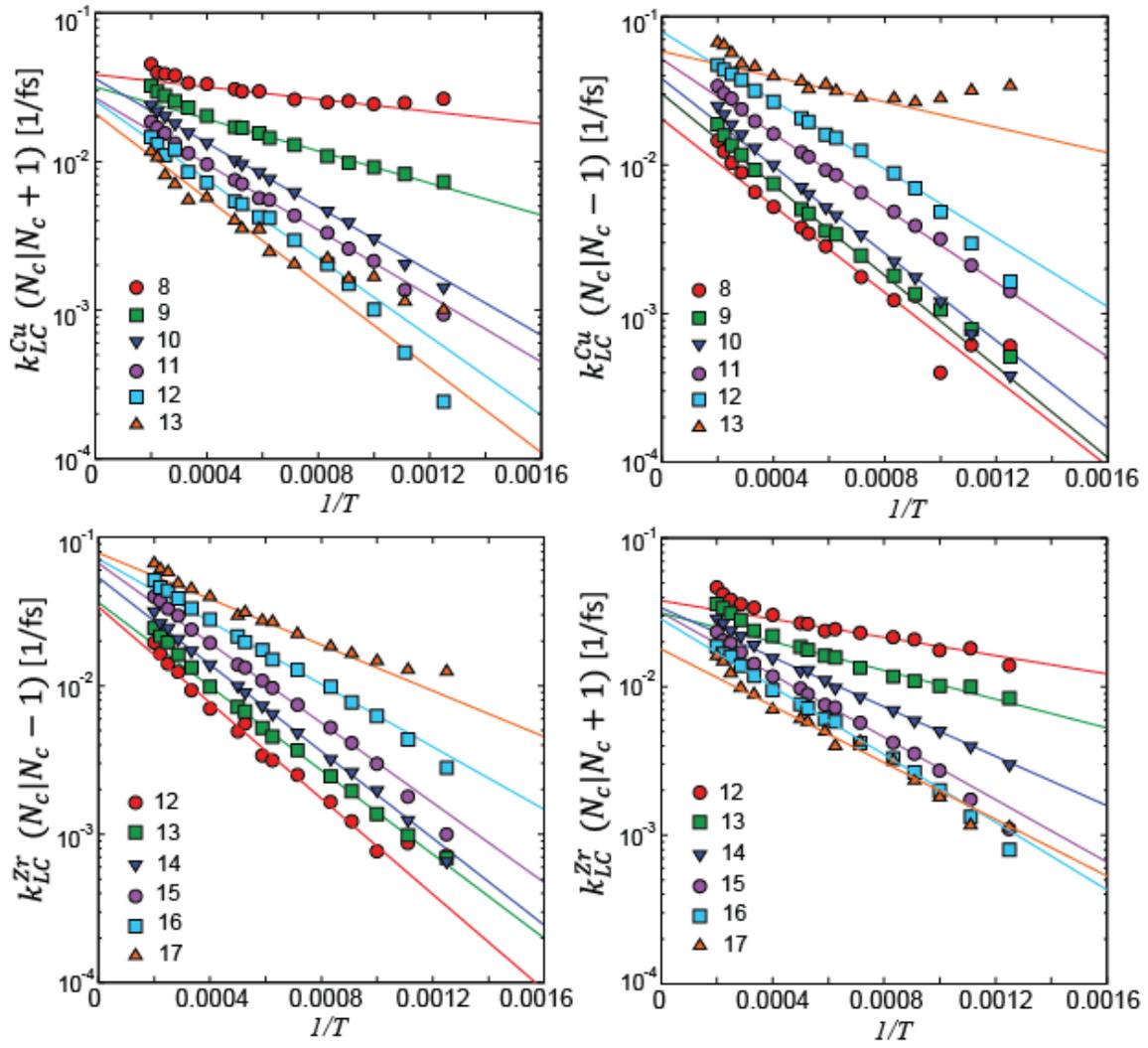

Fig. 7

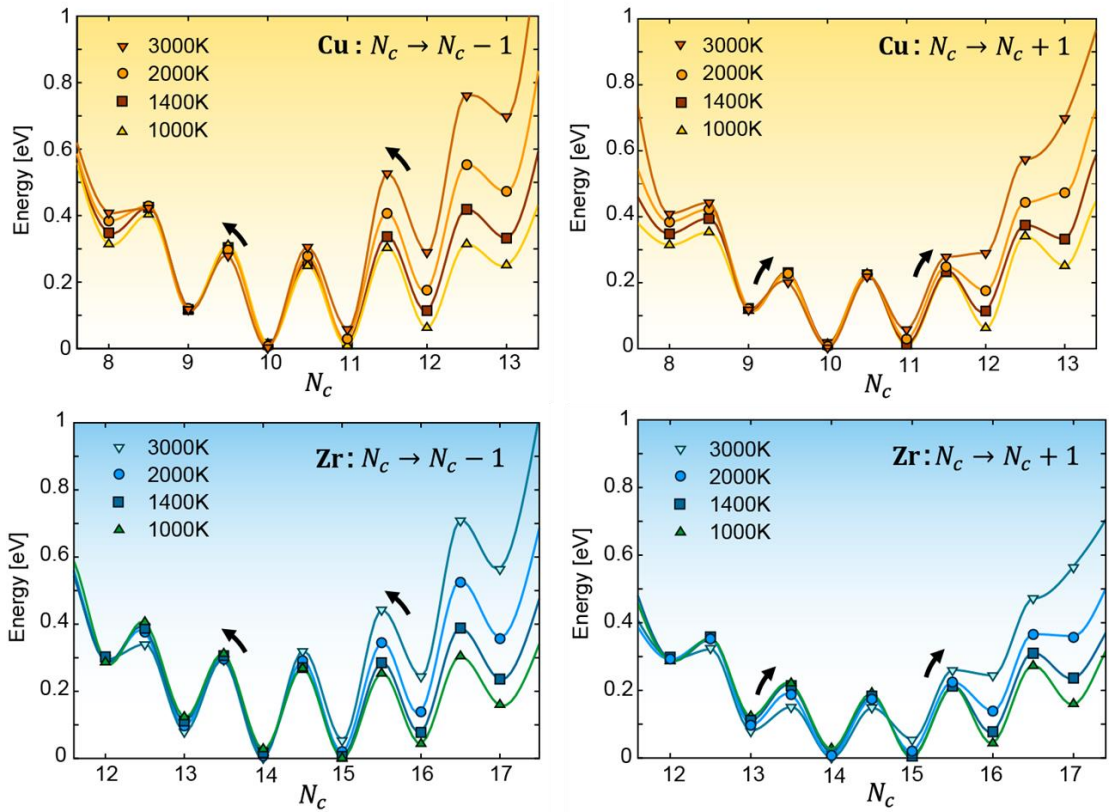

Fig. 8

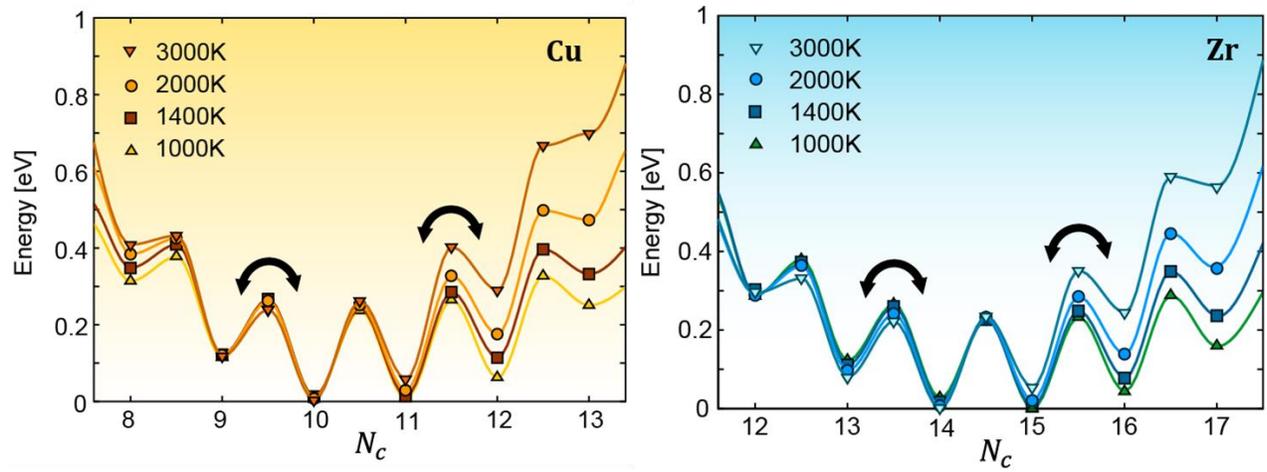

Fig. 9

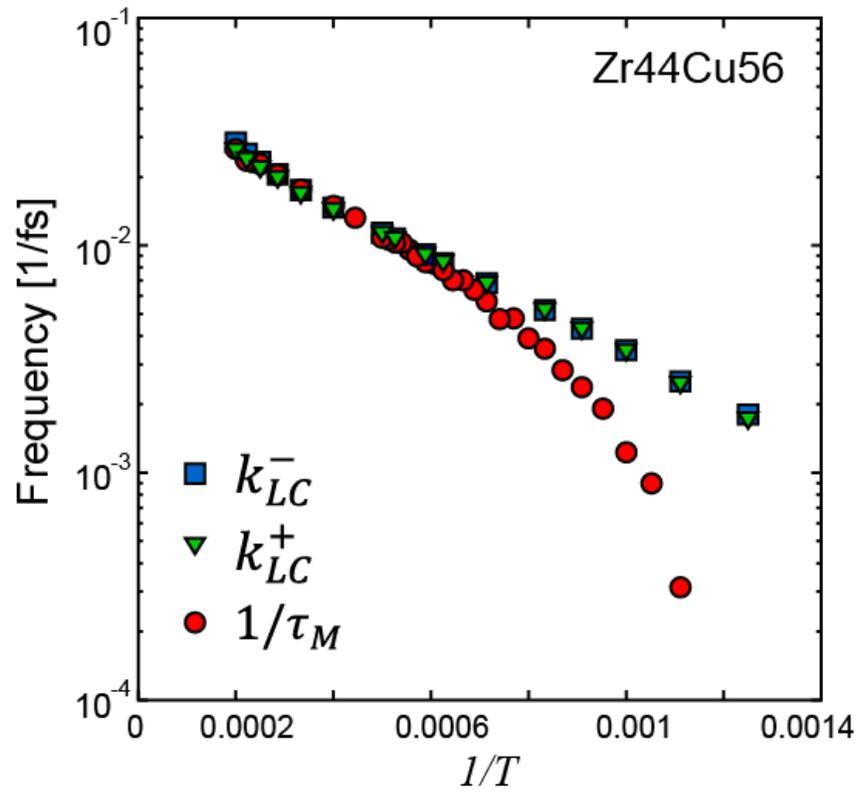

Fig. 10

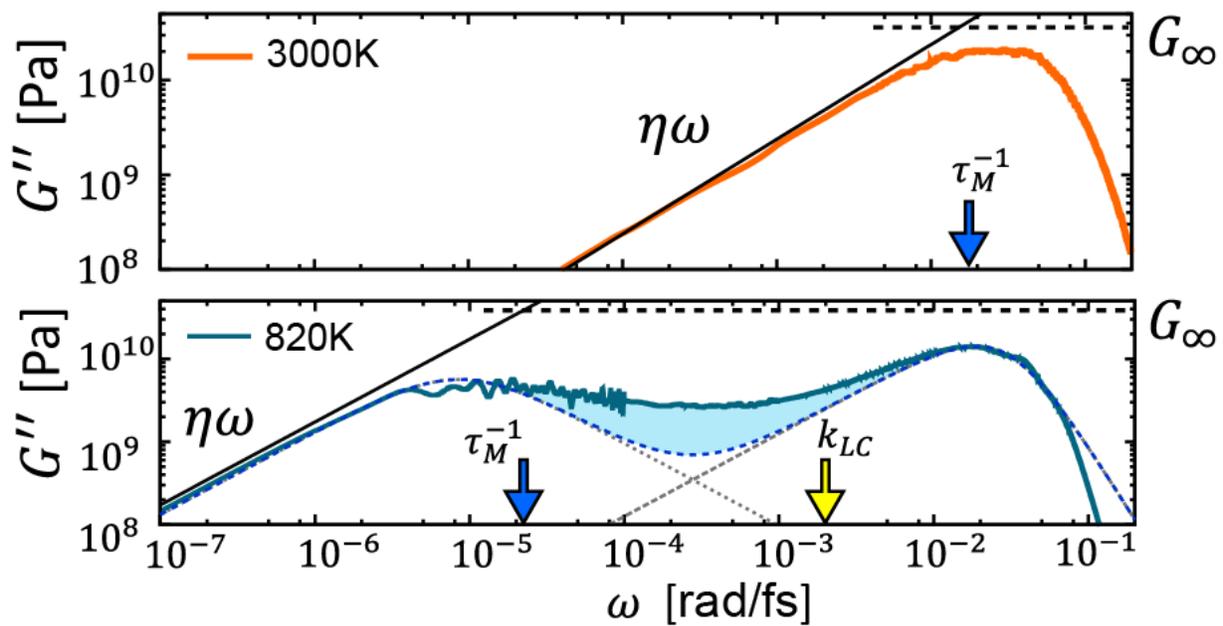

Fig. 11